\documentstyle[12pt,twoside,fleqn,espcrc1,psfig]{article}

\newcommand{\be}{\begin{equation}}
\newcommand{\ee}{\end{equation}}
\newcommand{\la}{\label}
\newcommand{\Z}{{\cal Z}}

\newcommand{\AmS}{{\protect\the\textfont2
  A\kern-.1667em\lower.5ex\hbox{M}\kern-.125emS}}

\title{Phase transition(s) in finite density QCD}

\author{R. Aloisio\address{Dipartimento di Fisica, Universit\'a dell'Aquila, 
        via Vetoio, 67100 L'Aquila, Italy}$^{,d}$,
	V. Azcoiti\address{Departamento de F\'isica Te\'orica,
	Facultad de Ciencias,Universidad de Zaragoza,50009 Zaragoza,Spain}
	G. Di Carlo\address{Istituto Nazionale di Fisica Nucleare,
	Laboratori Nazionali di Frascati, P.O.B. 13,
	00044 Frascati, Italy},
	A. Galante$^{a,c}$\thanks{Talk presented by A. Galante},
	A.F. Grillo\address{Istituto Nazionale di Fisica Nucleare,
	Laboratori Nazionali del Gran Sasso, 
	67010 Assergi, Italy}}

\begin{document}
\pagestyle{empty}

\maketitle

\begin{abstract}
The Grand Canonical formalism is generally used in numerical simulations of
finite density QCD since it allows free mobility in the chemical potential
$\mu$. 
We show that special care has to be used in extracting numerical
results to avoid dramatic rounding effects and spurious transition signals. 
If we analyze data correctly, with reasonable statistics, no signal of 
first order phase transition is present
and results using the Glasgow prescription
are practically coincident with the ones
obtained using the modulus of the fermionic determinant.
\end{abstract}

\section{INTRODUCTION}

Finite density QCD is a field where the lattice technology is still 
at an early stage.
Even if considerable progress has been achieved in the investigations of 
equation of state of QCD at finite temperature and zero chemical potential 
using the lattice approach,
the present situation of these studies at non zero density is not 
so satisfactory.

As well known, the complex nature of the determinant of the Dirac 
operator at finite chemical potential makes it impossible to use standard 
simulation algorithms based on positive-definite probability distribution
functions. 

Theoretical predictions in the lattice formulation are mainly limited to 
the infinite coupling limit and rely on some approximation ({\it e.g.} 
mean field).
In this limit they predict a strong first order saturation transition at a 
value for the chemical potential significantly smaller than one third of the 
baryon mass \cite{dagotto} (as one could naively expect considering the 
quarks confined inside hadrons but ignoring their binding energy).
The inclusion of some $\beta$ dependence in analytical calculations indicates
that the mean field critical density and ${1 \over 3} m_B$ converge
toward a common limit in the physically relevant scaling region \cite{bilic}.

Until recently only two 
results were available in numerical simulations, 
both in the strong coupling limit.
The first is based on a representation of the partition function as a system 
of monomers, dimers and baryonic loops (MDP) \cite{karsch}. 
The results show a clear first order signal at a critical chemical potential 
$\mu_c$ only slightly larger then the mean field prediction.

The second is based on an expansion in powers of the fugacity variable 
$z=e^\mu$: the 
Grand Canonical Partition Function formalism \cite{barbour}. 
The main advantage of this 
formalism is the free mobility in the chemical potential once we have all the 
coefficients of the expansion.
The non positivity of the integration measure can be overcome generating 
configurations accordingly to the Dirac determinant at zero chemical 
potential and defining $\Z=\langle \det\Delta(\mu)/\det\Delta(\mu=0)\rangle$
(also known as Glasgow algorithm).
The non positivity of the Dirac operator reflects in a non positivity of the
fugacity coefficients, at least for reasonable statistics. 
Nevertheless the complex coefficients can be used to compute the Lee-Yang 
zeros in the fugacity plane and extract some physical information.
Using this technique in the $\beta=0$ case, evidence has been found for 
a non saturation first order transition at a chemical potential in good 
agreement with the one extracted from MDP simulations \cite{barbour}.
Another interesting result of the Glasgow algorithm was the appearance 
of an early onset for 
the number density at a value $\mu_o$ going to zero in the chiral limit as in
simulations where the modulus of the fermionic determinant is considered.
In the latter case it can be explained as the effect of light baryonic states
formed from ordinary quarks and "conjugate" quarks interacting among
themselves through the complex conjugate of the Dirac operator \cite{davies}.
In QCD such states are certainly not present and the meaning of this
early onset has never been understood.

The aim of this paper is to show that the GCPF has some numerical subtleties
to be considered. Once rounding effects are under control the data  
show no signals of phase transition \cite{noi2}.
With reasonable statistics the only structure left is the early onset in the 
number density and no signals of first order phase transition survive:
numerical results are practically coincident with the theory defined using 
the modulus of the determinant. 

In the second section we introduce the GCPF formalism and show how
(wrong) numerical manipulations can simulate a first order phase transition
in the theory with quarks and conjugate quarks. The third section is devoted
to the study of numerical instabilities in the evaluation of coefficients of 
fugacity expansion. In the fourth the correct numerical analysis is reported 
together with conclusions.

\section{MODULUS OF THE DETERMINANT}

The contributions of the modulus of the Dirac determinant ($\Delta$)
and its phase $\phi_\Delta$ can be separated as
\be \la{1}
\Z = \Z_\| \langle e^{i\phi_\Delta} \rangle_\| \qquad\qquad\qquad\qquad
\qquad
\Z_\| = \int [dU] e^{-\beta S_g(U)} |\det\Delta(U,m,\mu)|
\ee
where $\langle e^{i\phi_\Delta} \rangle_\|$
accounts for the mean value of the cosine
of the phase computed with the probability distribution function of the pure
gauge theory times  $|\det\Delta|$.
It is clear that, in the thermodynamic limit, 
$\langle e^{i\phi_\Delta} \rangle_\|$ gives a net contribution
to the free energy density only in the case in which it falls off exponentially
with the lattice volume.
This is not what happens in $0+1$ dimensional QCD where the average of the
cosine of the phase is equal to $1$ \cite{lat97}.
To check if this is the case in $3+1$ dimensions we performed simulations 
of QCD at $\beta=0$ on different 
lattices using the modulus of the Dirac determinant \cite{noi1}.
We decided to generate configurations randomly after checking that the
effective fermionic action is a flat function of the pure gauge
energy around $E=1$.
The GCPF was used to reconstruct the coefficients of the fugacity
expansion from the eigenvalues of the propagator 
matrix $P$ \cite{gibbs}:
\be \la{P}
\det\Delta(m_q,\mu) = z^{3V}\det \left( P(m_q) - z^{-1}I\right) =
\sum_{n=-3V}^{3V} a_n(m_q) z^{n}
\ee
where $V$ is the lattice volume.

We considered the number density $n(\mu)$ as well as its derivative that, 
in the thermodynamic limit, is proportional to the radial distribution of 
Lee-Yang zeros in the complex fugacity plane $\rho(|e^\mu|)$:
\be \la{2}
\frac{d}{d\mu} n(\mu) = 4\pi e^{2\mu} \rho(|e^\mu|) .
\ee

\begin{figure}[!t]
\begin{minipage}[t]{75mm}
\psrotatefirst
\psfig{figure=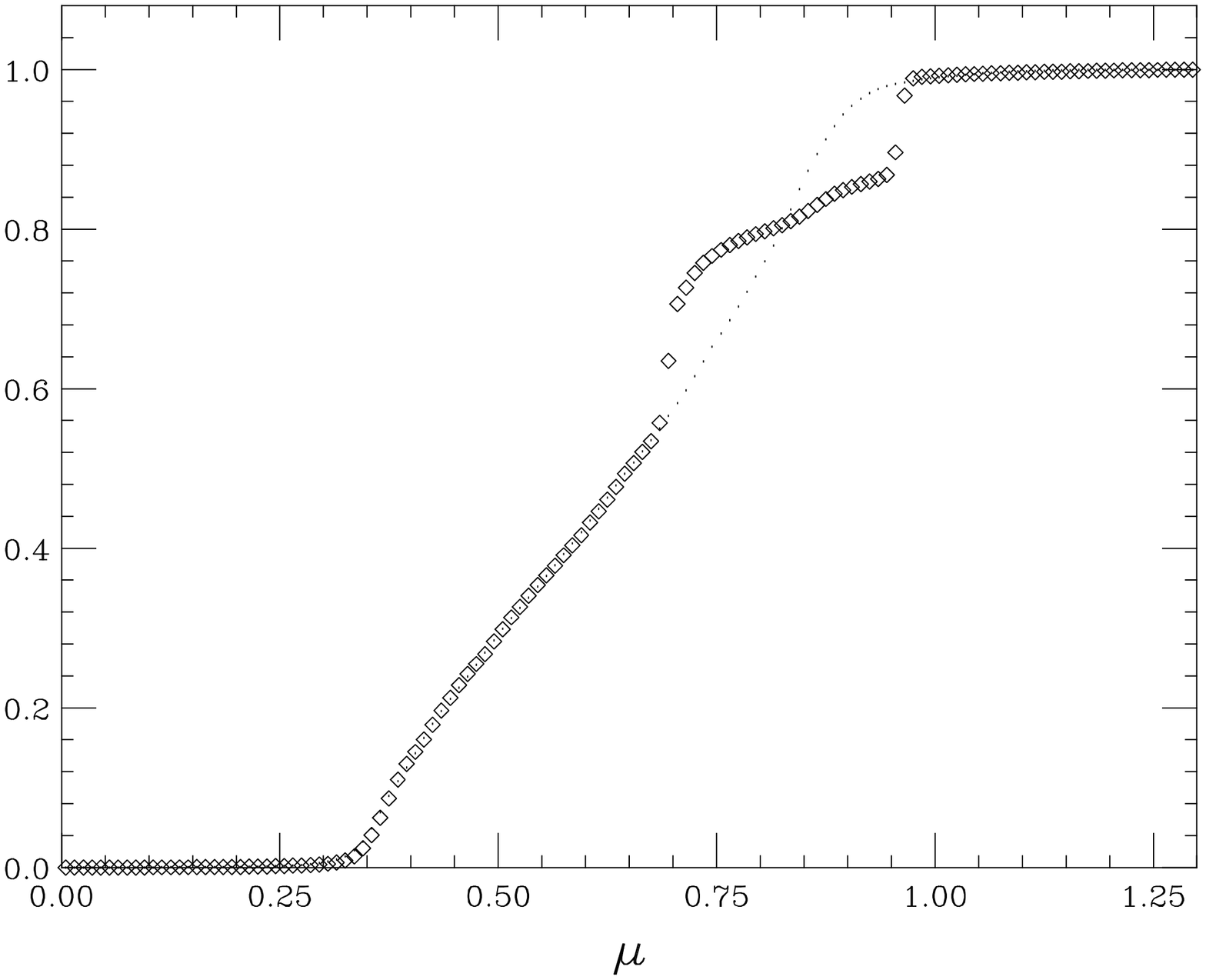,angle=90,width=210pt,
bbllx=45bp,bblly=80bp,bburx=510bp,bbury=650bp}
\caption{Number density in a $4^4$ (dots)
and $6^3\times 4$ (diamonds) at $m_q=0.1$ .}
\label{fig1}
\end{minipage}
\hspace{\fill}
\begin{minipage}[t]{75mm}
\psrotatefirst
\psfig{figure=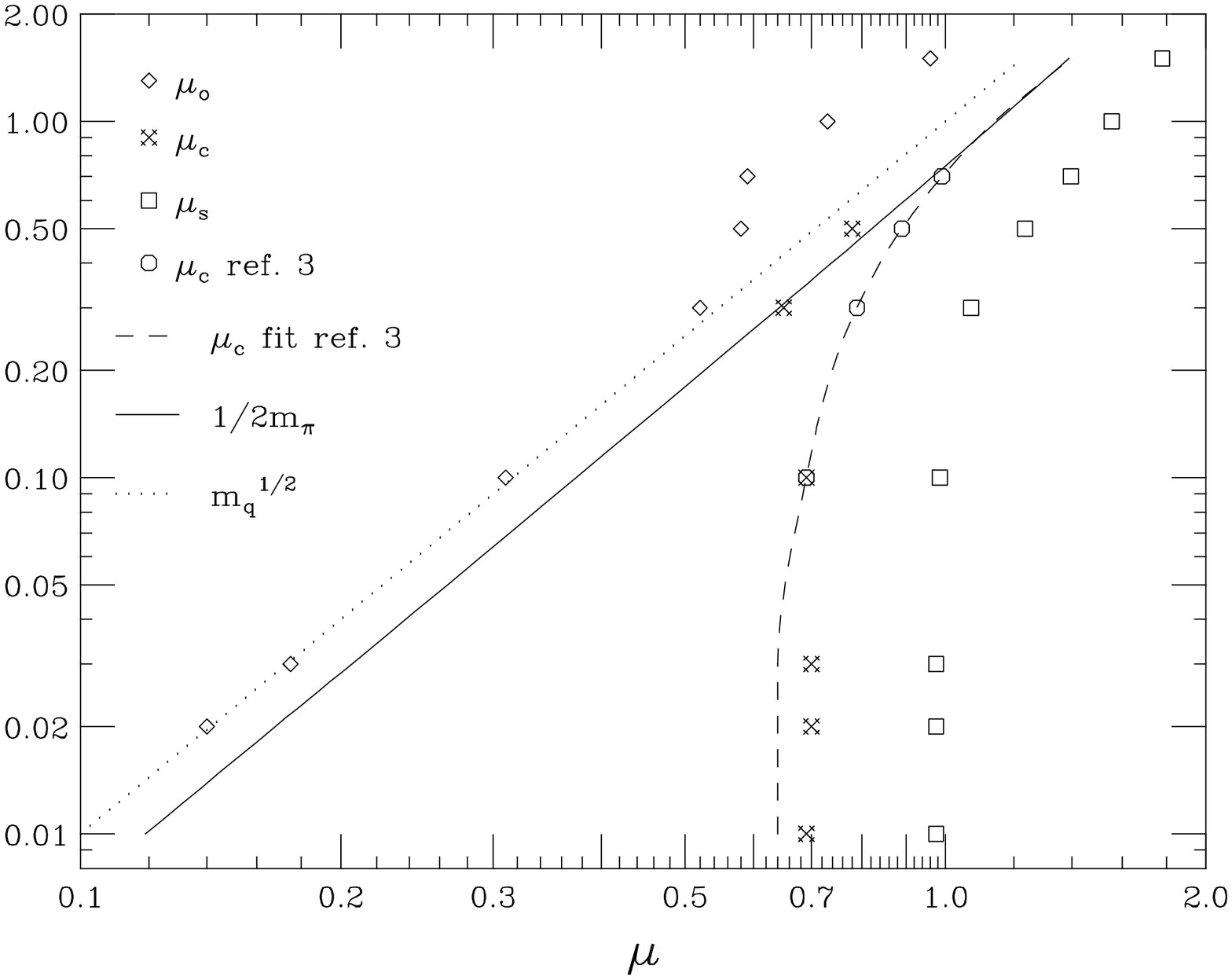,angle=90,width=210pt,
bbllx=45bp,bblly=80bp,bburx=510bp,bbury=650bp}
\caption{$\mu_o$, $\mu_c$, $\mu_s$ in the $(\mu,m_q)$ plane in a $6^3\times 4$
lattice.}
\label{fig2}
\end{minipage}
\end{figure}

The plots show results in some way contrary to common wisdom: 
if the system spatial extent is large
enough first order phase transition signals are evident (fig. \ref{fig1}).
The small $\mu$ behaviour is characterized by an onset threshold that
goes to zero in the chiral limit (fig. \ref{fig2}). For larger $\mu$ a clear
structure develops for $n(\mu_c)\simeq 0.6$ as well as at the saturation
point.
At small and intermediate masses, contrary to MDP results, the first order
critical point $\mu_c$ does not bring the system directly to saturation; 
nevertheless 
its position is in good agreement with what found using the MDP algorithm.
From $\mu=0$ up to $\mu_c$ there is also striking agreement with results
obtained using the Glasgow algorithm. For $\mu > \mu_c$ the agreement is
only qualitative but both methods show a saturation transition
at $\mu_s$.
On the contrary, in the quenched case, no signals of transition have been
found  indicating that the unquenching is essential to reproduce
the finite density physics \cite{barbour},\cite{noi1}.

We have also used the Glasgow
prescription ({\it i.e.} setting to zero the averaged $\bar a_n$ with negative 
real part) and calculated
the number density obtaining identical results.
The agreement between our calculations and the Glasgow algorithm is
surprising since the two models are believed to have completely different
baryonic spectrum.
Perhaps more surprising is the behaviour for large masses.
In this limit we can expect, on phenomenological grounds, a clear transition
signal since all the nucleons become confined in a single spatial
lattice site and do not interact each other.
This is not what is observed and $\mu_c$ disappears as soon as 
$m_q > 0.7$ (similar behaviour is reported also in \cite{barbour}). 

To have an independent check for this result we decided to diagonalize 
directly the Dirac matrix $\Delta(m_q=0)$ for various values of $\mu$:
for large masses a clear transition signal appears in agreement
with extrapolation of data from MDP (fig. \ref{fig3}) \cite{noi1}.

The discrepancy between results obtained from $P$ and $\Delta$ eigenvalues
points therefore to the existence of numerical problems in the evaluation of
thermodynamical quantities.

\section{ROUNDING EFFECTS}

There are three possible sources of numerical problems: 
$i)$ diagonalization of $\Delta$, $ii)$ diagonalization of $P$,
$iii)$ determination of the fugacity expansion coefficients $a_n$.

We used a standard NAG library routine to perform the diagonalization of
$\Delta$ and $P$.
Using the eigenvalues of the two matrices (constructed from the same gauge
configuration) we verified that the equality between $\det\Delta$
and $z^{3V}\det(P-z^{-1}I)$ (the latter evaluated directly from $P$
eigenvalues) holds in the whole range of $\mu$ and masses we considered.
At $\mu=0$, we also found a perfect agreement between the 
$\Delta$ eigenvalues computed with this routine with the ones obtained using a 
standard Lanczos algorithm.

We conclude that the diagonalization procedure is stable for any value of 
$m_q$ and $\mu$ and that the numerical problems
can only be due to the manipulations necessary to go from the $P$ 
eigenvalues to the GCPF coefficients $a_n$.
The only feasible way to perform this calculation is to adopt a 
standard recursion method. We use the
first $n$ eigenvalues to calculate the coefficients of a
polynomial $Q_n$ of degree $n$:
\be \la{coeff}
a_k^{(n)} = a_{k-1}^{(n-1)} - \lambda_n a_k^{(n-1)} \qquad\qquad k\ge 1
\ee
where $a_k^{(n-1)}$ is the coefficient of order $k$ of $Q_{n-1}$ and
$\{\lambda_i\}$ are the eigenvalues.

This algorithm has problems similar to the deflating technique for finding
the zeros of large polynomials \cite{sharpe}. 
We can get the right answer only if we are
able to reproduce the (approximate) symmetries of the correct result at
each intermediate step of calculation. In this case the coefficients
never grow too much and we do not have to rely on cancellation of 
large numbers to get the right answer.

To clarify this point it is useful to consider a simple case: a single
configuration whose $\{\lambda_i\}$ are uniformly distributed on two 
circles of radius $\rho$ and $\rho^{-1}$ (to fulfill the 
$\lambda$, $1/\lambda^*$ symmetry).
The fugacity polynomial contains only three non vanishing terms:
$a_{\pm 3V} = 1$ and $a_0 = \rho^V + \rho^{-V}$.
In the infinite volume limit this corresponds to a first order 
saturation transition with the number density that can be approximated by a
Heaviside $\theta$ function.

If we use (\ref{coeff}) to calculate the coefficients we can get different
results depending on the ordering of the eigenvalues \cite{noi2}. 
If the eigenvalues are phase-ordered, at any intermediate step we calculate
the coefficients of a polynomial whose zeros lay on two arcs of circle of
increasing length.
These coefficients are non zero and of order $O(\rho^{n})$: in this case
rounding propagation easily prevents to obtain the correct answer.
This happens already for relatively small $V$
and forbids the symmetries to be realized in the final results. 
If we use randomly ordered eigenvalues we modify this scenario since the 
symmetries are (almost) enforced at each intermediate step as well as in 
the final result \cite{noi2}:
the coefficients of $Q_n$ never grow too much and rounding
effects are better under control.

The eigenvalues from the diagonalization routine 
can be (partly) phase ordered (and this is what often 
happens with our NAG routine): they can not be 
considered a good input for the routine that evaluates the coefficients.
Rounding effects become more and more probable as the volume increases
since in this case the coefficients are bigger and a smaller fraction of
ordered $\{\lambda_i\}$ can suffice to create problems.

\begin{figure}[!t]
\begin{minipage}[t]{75mm}
\psrotatefirst
\psfig{figure=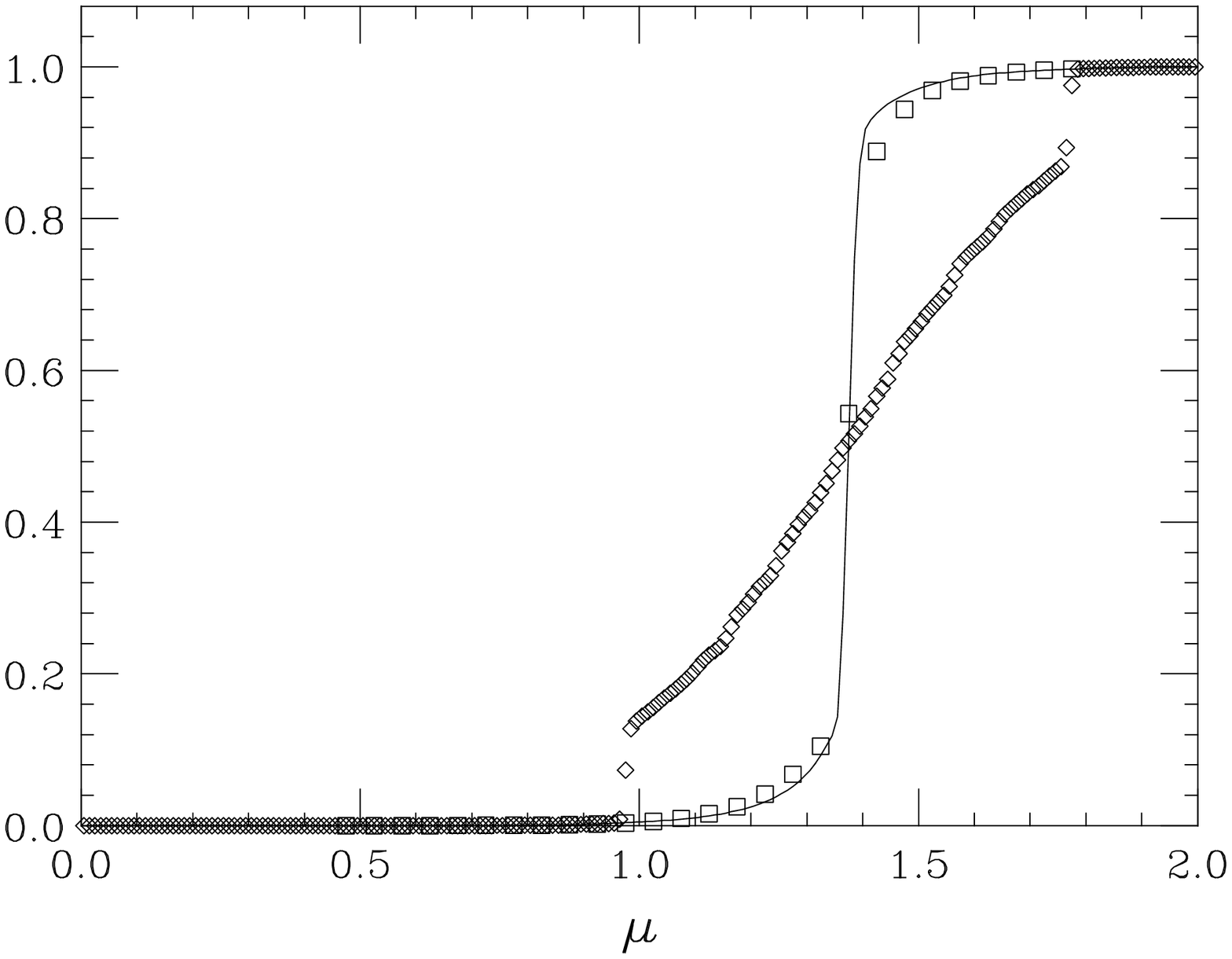,angle=90,width=210pt,
bbllx=45bp,bblly=80bp,bburx=510bp,bbury=650bp}
\caption{Number density in a $6^3\times 4$ lattice at $\beta=0$ and $m_q=1.5$
with shuffled (solid line) and unshuffled (diamonds) eigenvalues;
the same quantity for a $4^4$ lattice from the $\Delta$ eigenvalues (squares).}
\label{fig3}
\end{minipage}
\hspace{\fill}
\begin{minipage}[t]{75mm}
\psrotatefirst
\psfig{figure=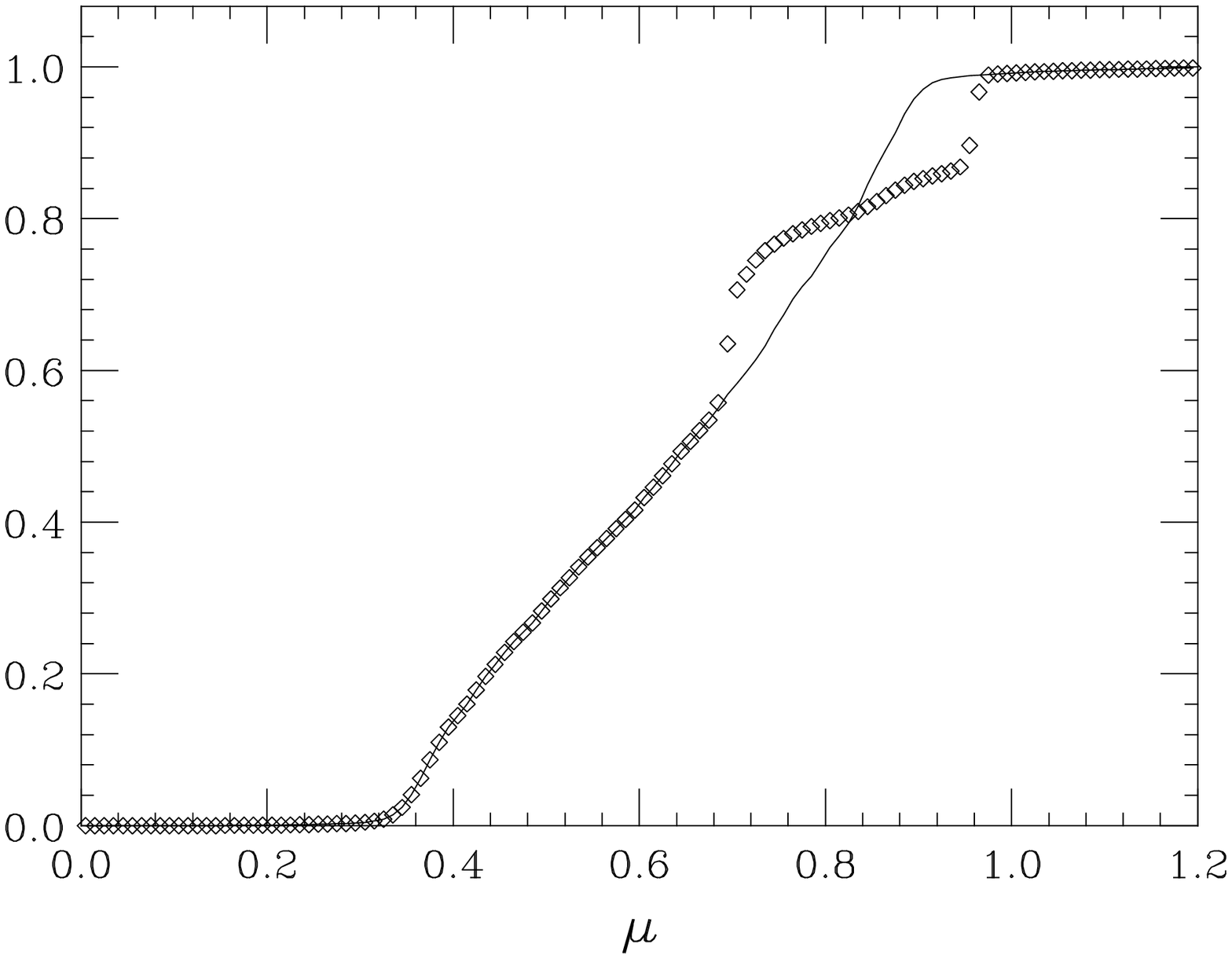,angle=90,width=210pt,
bbllx=45bp,bblly=80bp,bburx=510bp,bbury=650bp}
\caption{Number density in a $6^3\times 4$ lattice at $\beta=0$ and $m_q=0.1$
using shuffled (solid line) and unshuffled (diamonds) eigenvalues.}
\label{fig4}
\end{minipage}
\end{figure}

For actual configurations the eigenvalues are distributed in two (almost)
circular strips in the complex plane. In this case we have to pay attention
to the ordering of their modulus too.
The shuffling procedure solve both problems since at each intermediate step
the zeros of $P_n$ have, in the complex plane, the same distribution as
the $\{\lambda_i\}$. 
When we do it we get from the coefficients results
indistinguishable from the ones obtained  computing the fermionic determinant
directly from the propagator matrix eigenvalues.
In fig. \ref{fig3} is shown that once we solve the numerical problems we 
can reconcile results from $P$ and $\Delta$ eigenvalues. 

\section{STRONG COUPLING RESULTS}

After we have removed the numerical artifacts we can reconsider
the strong coupling results at small quark masses. 
The plot in fig. \ref{fig4} shows that data, if correctly analyzed,
give no evidence of first order transition (the plot refers to data obtained
using the Glasgow prescription but, using the modulus, we get the same results
\cite{noi2}).
No signal is present in the position indicated from MDP results and
the only structure left is the onset $\mu_o$, once again consistent with
the presence in the spectrum of a light nucleon made of an ordinary quark and 
a conjugate quark \cite{davies}.

This result can be easily understood when we realize that
between the onset $\mu_o$ and the saturation point the 
distribution of the phase of the determinant becomes indistinguishable
from a flat one: the averaged determinant is
no longer a real and positive quantity as it should be.
The free energy density defined through the average of the modulus differs
from the complete one by means of the phase contribution (\ref{1}).
With $N$ independent configurations the best we can do is
to evaluate $\langle e^{i\phi_\Delta} \rangle_\|$ with an error of the
order of $O(\frac{1}{\sqrt{N}})$. 
If the contribution of the phase is relevant it has to be exponentially 
small with the system volume but we can not appreciate its contribution 
unless $N = O(e^V)$.

Except for the first, central and last coefficient (constrained to
be real and positive from eigenvalues symmetries) the same scenario is valid
for all the coefficients in the fugacity expansion.
The coefficients are canonical partition functions (at fixed number density)
and, in perfect analogy with the  grand canonical partition function, 
the modulus of their average practically coincide with the average of 
their modulus.

From this considerations we can understand why the Glasgow prescription
gives the same result as the modulus: unless we have a huge
statistics (exponential with the system volume) all the possible 
definitions for a real and positive $\Z$
point toward the theory with the modulus of the Dirac determinant \cite{noi2}
(see \cite{barbour2} for a high statistics study of $2^4$).
The MDP algorithm is not plagued from the same problems since 
one first integrates over the gauge fields and then over the matter field.
In some sense in this case one avoids the sign problem summing
over all possible gauge fields exactly.
The main conclusion is that the moderate optimism that was inspired by the
GCPF calculations in the last years has to be considered ill-founded.

\end{document}